\documentclass[english,a4paper,prb,twocolumn]{revtex4}
\usepackage[T1]{fontenc}
\usepackage[latin1]{inputenc}
\usepackage{amsmath}
\usepackage{graphicx}
\usepackage{amssymb}

\makeatletter

\usepackage{calc}

\usepackage{xspace}

\usepackage{bm}

\usepackage{ifthen}

\newcounter{hours}
\newcounter{minutes}
\newcommand{\printtime}{
\setcounter{hours}{\time/60}
\setcounter{minutes}{\time-\value{hours}*60}
\ifthenelse{\value{hours} < 10 }{0\thehours:}{\thehours:}
\ifthenelse{\value{minutes} < 10}{\negthickspace0\theminutes}{\negthickspace\theminutes}
}

\usepackage{babel}
\makeatother
\begin{document}

\title{Spin-orbit-assisted electron-phonon interaction and the magnetophonon
resonance in semiconductor quantum wells}

\author{D. S. L. Abergel and Vladimir I. Fal'ko}

\affiliation{Physics Department, Lancaster University, Lancaster, LA1 4YB, UK}

\begin{abstract}
We introduce a spin-orbit assisted electron-phonon (e-ph) coupling
mechanism for carriers in semiconductor quantum wells and predict
qualitatively the form and shape of anticrossings between the cyclotron
resonance (CR) and optical (LO and TO) phonons. Since this e-ph interaction
involves the electron spin it makes the predicted TO-CR anticrossing
dependent on the filling factor and therefore on the polarization
of the 2D electron gas in the quantum well.
\end{abstract}
\maketitle
Recently, the fabrication of very thin semiconductor films with embedded
quantum wells (QW), \textit{e.g.}, in GaAs/AlGaAs structures has
made it possible to perform cyclotron resonance (CR) experiments in
the frequency range which covers the restrahlung band in the material\cite{bib:martinez}.
As a result, the regime of the CR crossing the longitudinal (LO) or
transverse (TO) optical phonon - where the magnetoplasmon in the two-dimensional
electron gas (2DEG) is in resonance with these optical lattice vibrations
- became feasible to study, subject to the coupling between these
modes.

Near the crossings, the cyclotron resonance and the optical phonon
modes may mix and form anticrossings. However for $q=0$ there can
be no direct CR-optical phonon coupling due to the lattice symmetry
of III-V semiconductors\cite{bib:martinez,bib:Bir-book,bib:faugeras-prl92,bib:klimin-prb68}.
If the incident radiation and the magnetic field in the experiment
are orientated perpendicularly to the plane of the 2DEG, there is
no coupling allowed in a QW as the CR crosses the LO phonon mode because
there is no mixing of the electric field of the in-plane motion of
the electrons and the out-of-plane motion of the lattice. There is
also no coupling when the CR crosses the TO mode because this phonon
does not carry any electric field and the deformation potential with
$q=0$ does not allow for the emission or absorption of TO phonons
by electrons (it only allows for e-ph coupling in the second order
of the lattice displacement, that is, electron-phonon scattering).

In this Communication, we propose a mechanism of electron-phonon interaction
which does permit the mixing of the CR and phonon modes\cite{bib:ff-range}.
This mechanism is assisted by the spin-orbit (SO) coupling in the
atomic shells of the material and involves the electron spin in the
interaction process. We find that in a perpendicular magnetic field,
the CR will couple to the TO phonon, with the strongest coupling occuring
when the electron gas is fully spin-polarized (for example, at $\nu=1$)
and reducing to zero when the electron gas is completely un-spin-polarized
(\textit{e.g.} at $\nu=2$).

We model the quantum well containing the 2DEG as a thin slab of polar
semiconductor mounted on a non-polar substrate. Electrons are confined
in a QW which is embedded in a matrix of a material which has not
only a different bandgap but also a non-overlapping phonon density
of states. For a thin slab there are two phonon modes polarised in
the plane of the slab with the frequency $\omega_{x,y}(q\to0)=\omega_{TO}$
and one mode polarised perpendicularly to the slab with $\omega_{z}(q\to0)=\omega_{LO}$.

For semiconductors with a zinc-blende lattice structure we write down\cite{bib:Falko-prb,bib:Goerbig-arXiv}
the SO-assisted (Rashba-type\cite{bib:rashba}) electron interaction
with lattice displacements as\cite{bib:intham} \begin{equation}
\mathcal{H}_{e-ph}=\frac{\alpha}{2}\boldsymbol{\sigma}\cdot(\mathbf{p\times w+w\times p}).\label{eq:V_def}\end{equation}
This term is invariant with respect to the transformations from the
symmetry group of the zinc-blende type lattice, $T_{d}$. The vector
$\mathbf{w}$ represents the relative displacement of the two sublattices
with respect each other, which acts to deform the atomic bonds. This
in turn changes the hopping integrals of electrons between SO-mixed
orbitals and provides the mechanism for the coupling in Eq. \eqref{eq:V_def},
parameterized using a material-dependent phenomenological constant
$\alpha$.

For optical phonons confined to a $N$-atomic-layer thick QW, the
sublattice displacement field $\mathbf{w}(\mathbf{r})=\mathbf{w}(x,y)\beta_{k}(z)$
represents a standing wave in the slab\cite{bib:phonon} with $\beta_{k}(z)=\sqrt{2/N}\sin(\pi kz/Na)$.
The second quantised phonon field operator can be written as \begin{equation}
\mathbf{w}=\sum_{\mathbf{q},\kappa}\sqrt{\tfrac{\hbar a^{2}}{2L_{x}L_{y}m_{\kappa}\omega_{\kappa}}}e^{i\mathbf{q\cdot\mathbf{x}}/\hbar}\beta(z)\mathbf{l}_{\mathbf{q},\kappa}(b_{\mathbf{q},\kappa}+b_{-\mathbf{q},\kappa}^{\dagger})\label{eqn:w_full}\end{equation}
 where $\kappa=x,y,z$ denotes the phonon modes, $\mathbf{l}_{\mathbf{q},\kappa}$
is the unit polarisation vector and $b$ and $b^{\dagger}$ stand
for annihilation and creation operators of phonons. The reduced mass
of the atoms in the unit cell and frequency of the phonon modes are
$m_{\kappa}$ and $\omega_{\kappa}$ respectively, $a$ is the lattice
constant and $L_{x,y}$ are the lateral sample size.

The high magnetic field CR can be discussed in terms of electron transitions
between Landau levels upon absorption of a far-infrared photon. The
Landau levels involved in such transitions depend on the filling factor
and the sign of the conduction band $g$-factor. Below we study a
2DEG with $\nu\leq2$ where electrons occupy only the lowest ($n=0$)
Landau level\cite{bib:ff-range}. In our analysis the spin polarisation
of the 2DEG matters because the electron spin is involved in the interaction.
Therefore we might expect the qualitative features of the coupling
to differ for $\nu=1$ (a fully spin-polarized 2DEG) and $\nu=2$
(an un-spin-polarized 2DEG). The inter-Landau level transtitions for
the QHE states at $\nu\lesssim2$ of electrons can be described using
the magnetoexciton operators which define coherent excitations of
electrons from the $n=0$ to the $n=1$ Landau level \cite{bib:bychkov,bib:kallin}:
\begin{equation}
\Psi_{\alpha,\alpha^{\prime}}^{\dagger}(\mathbf{Q})=\sum_{p}\frac{e^{ipQ_{y}}}{\sqrt{\mathcal{N}}}a_{1,\alpha,p}^{\dagger}a_{0,\alpha^{\prime},p-Q_{x}}.\end{equation}
 The indices $\alpha,\alpha^{\prime}$ represent the spin of final
and initial states respectively, $\mathcal{N}$ is the number of single
electron states in each Landau level and $a,(a^{\dagger})$ are single
electron annihilation (creation) operators.

\begin{figure}[tb]
\centering \includegraphics{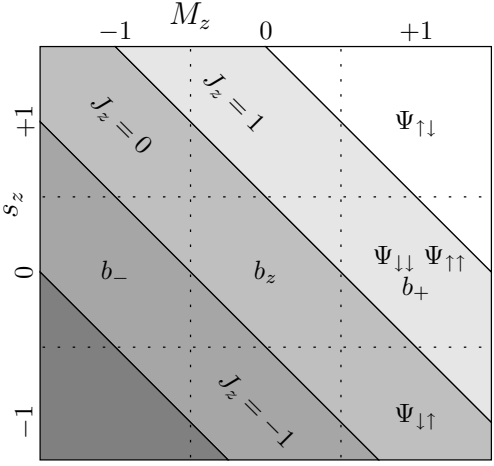}

\caption{This figure shows the quantum numbers of each of the primary modes
in the system. When the magnetic field is perpendicular to the 2DEG,
coupling is allowed only between modes which conserve $J_{z}=M_{z}+s_{z}$,
shown in regions of the same colour. }

\label{fig:JMS-table} 
\end{figure}

For $\nu=2$ the ground state has electrons completely filling both
the up- and down-spin levels. The low-energy excitations are then
the spinless (CR) mode $S=(\Psi_{\uparrow\uparrow}+\Psi_{\downarrow\downarrow})/\sqrt{2}$
which has $s_{z}=0$ and energy $\hbar\omega_{c}$ and three triplet
modes which carry spin $s=1$: $\Psi_{\uparrow\downarrow}$ which
has $s_{z}=+1$ and energy $\hbar\omega_{c}+\epsilon_{z}$ (the Zeeman
energy $\epsilon_{z}=g\mu_{B}B$ can be positive or negative depending
on the sign of the $g$-factor, which we do not specify); $\Psi_{\downarrow\uparrow}$
which has $s_{z}=-1$ and energy $\hbar\omega_{c}-\epsilon_{z}$;
and $T=(\Psi_{\uparrow\uparrow}-\Psi_{\downarrow\downarrow})/\sqrt{2}$
which has $s_{z}=0$ and energy $\hbar\omega_{c}$.

When $\nu=1$ the modes which can be excited in the 2DEG with frequency
close to $\omega_{c}$ are defined by the sign of the $g$-factor.
If this is negative then the 2DEG is a fully polarised system of up-spin
electrons giving rise to excitations $\Psi_{\uparrow\uparrow}$ (which
corresponds to the CR and has $s_{z}=0$ and energy $\hbar\omega_{c}$)
and $\Psi_{\downarrow\uparrow}$ (which has $s_{z}=-1$ and energy
$\hbar\omega_{c}+\epsilon_{z}$). In this case the Zeeman energy contains
a contribution from the electron-electron exchange interaction. If
the $g$ -factor is positive then the 2DEG is fully polarised spin-down
and at $\omega\approx\omega_{c}$ the excitations are $\Psi_{\downarrow\downarrow}$
(the CR which carries $s_{z}=0$) and $\Psi_{\uparrow\downarrow}$
(which has $s_{z}=+1$). All electronic excitations have $M_{z}=+1$.

To analyse the fine structure of the CR at the CR-LO and CR-TO crossings
we determine the effective coupling between electron inter-Landau
level excitons and the lattice vibrations. We represent the wave functions
of electrons in the QW subjected to a high magnetic field in the form
$\Phi(\mathbf{r})=\psi_{n,p}(y/\lambda_{B})\varphi_{0}(z)$ adapted
to the Landau guage with vector potential $\mathbf{A}=-B_{z}y\mathbf{l}_{x}$.
The function $\varphi_{0}(z)=\sqrt{2/Na}\sin(\pi z/Na)$ describes
the form of the electron wave function in the QW and the Landau level
states $\psi_{n,p}$ are related to each other by raising and lowering
operators $\pi=-\lambda_{B}/\sqrt{2}\hbar(p_{x}-ip_{y})$ and $\pi^{\dagger}=-\lambda_{B}/\sqrt{2}\hbar(p_{x}+ip_{y})$
such that $\pi^{\dagger}\psi_{n,p}=\sqrt{n+1}\psi_{n+1,p}$, $\pi\psi_{n,p}=\sqrt{n}\psi_{n-1,p}$
and $\pi\psi_{0,p}=0$. Since each electron in the 2DEG covers many
($N$) atomic layers, the effective coupling of an electron is\begin{equation}
Z_{p}=\frac{2\pi\xi_{k}\hbar a}{\lambda_{B}}\sqrt{\frac{2\hbar}{m_{p}\omega_{p}}},\end{equation}
which is reduced by the factor \[
\xi_{k}=\frac{\alpha}{2}\int_{0}^{Na}\beta_{k}(z)|\varphi_{0}(z)|^{2}\, dz=\begin{cases}
0 & \text{even $k$,}\\
\frac{4\alpha}{k(4-k^{2})\pi}\sqrt{\frac{2}{N}} & \text{odd $k$.}\end{cases}\]
The coupling is strongest with the lowest mode $k=1$ which will be
the only mode we discuss in the following analysis. The subscript
$p=LO,TO$ denotes the phonon branch.

When $\mathbf{B}=B\mathbf{l}_{z}$, the rotational symmetry about
the $z$ axis of both the Hamiltonian and the interaction makes the
projection $J_{z}=M_{z}+s_{z}$ of the total angular momentum carried
by an excitation an exact quantum number. As a result only modes which
have the same value of $J_{z}$ can mix by the interaction in Eq.
(\ref{eq:V_def}), as illustrated in Fig. \ref{fig:JMS-table} where
each excitation has been placed according to its orbital momentum
and spin. The electron and phonon modes which can mix must belong
to the same constant-$J_{z}$ `diagonal' highlighted using shaded
stripes. To define TO phonon modes with fixed angular momentum near
the $\Gamma$-point (center) of the Billouin zone we introduce the
linear combinations of degenerate $x$- and $y$-polarised vibrations\cite{bib:noq,bib:Goerbig-arXiv}
\begin{equation}
b_{\pm}=\left(b_{y}\pm ib_{x}\right)/\sqrt{2}.\end{equation}
 Then $b_{+}$ becomes the annihilation operator for the phonon carrying
$M_{z}=+1$ and $b_{-}$ for $M_{z}=-1$. For completeness we mention
that the $z$-polarised vibration $b_{z}$ (with LO phonon frequency)
has $M_{z}=0$ and all phonon modes have $s=0$. For example, the
selection rule for the mode mixing illustrated in Fig. \ref{fig:JMS-table}
shows that the phonon mode $b_{-}$ and the spin-flip mode $\Psi_{\uparrow\downarrow}$
cannot mix with any other modes but that $b_{+}$ may couple only
to $\Psi_{\uparrow\uparrow}$ or $\Psi_{\downarrow\downarrow}$. Below
we consider seperately the couplings at filling factor $\nu=1$ (for
positive and negative $g$) and for filling factor $\nu=2$ since
each of these systems has its own set of magneto-excitons.

The fine structure of the spectrum of modes $\Omega_{ep}^{\pm}$ near
a crossing can be described by the equation \begin{equation}
\hbar\Omega_{ep}^{\pm}=(\hbar\omega_{e}+\hbar\omega_{p})/2\pm\sqrt{(\hbar\omega_{e}-\hbar\omega_{p})^{2}/4+\gamma Z_{p}^{2}},\label{eq:split-energies}\end{equation}
found by diagonalising the appropriate Hamiltonian directly. The subscripts
$e$ and $p=LO,TO$ identify the electronic and phonon modes involved
in the coupling respectively. The numerical parameter $\gamma$ is
filling-factor-dependent and should be specified for each individual
individual pair of modes. We calculate the size of the minimum splitting
as $\hbar(\Omega_{ep}^{+}-\Omega_{ep}^{-})$ and label it by $\delta_{p}$
for optically active electronic modes, $\tilde{\delta}_{p}^{0}$ for
passive modes with $s_{z}=0$, and $\tilde{\delta}_{p}^{s}$ for passive
modes with $|s_{z}|=1$.

\begin{figure}
\centering\includegraphics[bb=152bp 540bp 389bp 668bp,clip]{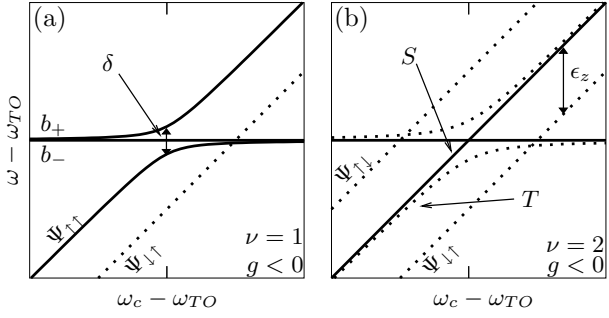}

\caption{Anticrossings of electronic modes and the TO phonon. Throughout the
solid lines denote optically active modes (the CR) and phonon modes
while dotted lines represent the optically passive spinful electronic
modes. (a) When the CR crosses the TO mode at $\nu=1$ there is a
splitting of magnitude $\delta_{TO}=2Z_{TO}$. (b) When $\nu=2$ the
CR is decoupled from the TO phonons ($\delta_{TO}=0$), but the spinful
magnetoexciton $T$ mixes with $b_{+}$ and has a splitting of $\tilde{\delta}_{TO}^{0}=2\sqrt{2}Z_{TO}$.
The tick marks denote the point $\omega_{c}=\omega_{TO}$.\label{fig:TOsplitting}}
\end{figure}

\begin{figure}
\centering \includegraphics[bb=150bp 430bp 389bp 669bp,clip]{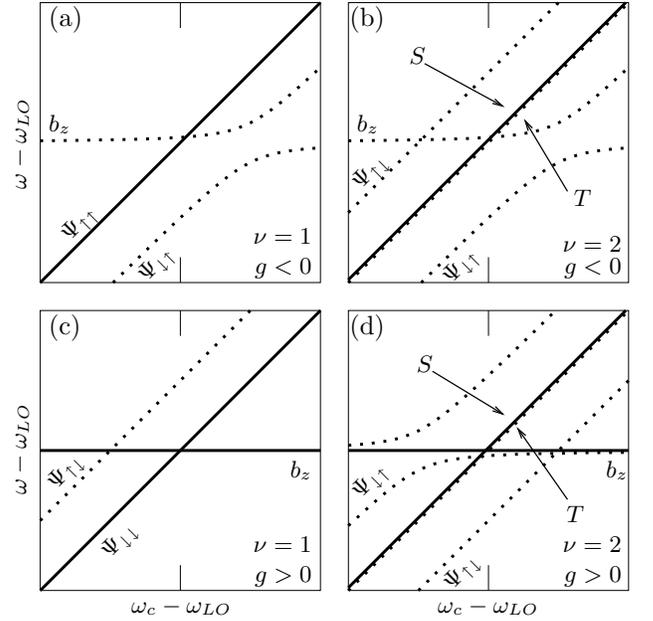}

\caption{Crossings in the region of the LO phonon. Solid lines denote optically
active electronic modes (that is, the CR) and the phonon modes; dotted
lines represent spinful electronic modes. (a) When $\nu=1$ and $g<0$
the CR is decoupled from the LO phonon, but there is a splitting of
magnitude $\tilde{\delta}_{LO}^{s}=2\sqrt{2}Z_{LO}$ between the LO
phonon and the spinful $\Psi_{\downarrow\uparrow}$ mode. (b) For
$\nu=2$ and $g<0$ the CR is decoupled from the LO phonon, but the
mode $\Psi_{\downarrow\downarrow}$ does split with magnitude $\tilde{\delta}_{LO}^{s}=2\sqrt{2}Z_{LO}$.
(c) When $\nu=1$ and $g>0$ there is no coupling of any modes. (d)
For $\nu=2$ and $g>0$, the coupled spinful mode splits from the
LO phonon with magnitude $\tilde{\delta}_{LO}^{s}=2\sqrt{2}Z_{LO}$.
Tick marks on the horizonal axis show where $\omega_{c}=\omega_{LO}$.\label{fig:LOsplitting} }
\end{figure}

For $\nu=1$ and $g<0$ the electrons fill the up-spin states and
therefore the magnetoplasmons in this configuration are the optically
passive $\Psi_{\downarrow\uparrow}$ and the CR mode $\Psi_{\uparrow\uparrow}$.
The Hamiltonian describing possible couplings of the electron and
phonon modes generated by the microscopic mechanism in Eq. \eqref{eq:V_def}
can be written down as \begin{multline}
\hat{\mathcal{H}}_{\nu=1,g<0}=\left[\Psi_{\uparrow\uparrow}^{\dagger},\hat{b}_{+}^{\dagger},\Psi_{\downarrow\uparrow}^{\dagger},b_{z}^{\dagger}\right]\times\\
\times\begin{bmatrix}\hbar\omega_{c} & Z_{TO} & 0 & 0\\
Z_{TO} & \hbar\omega_{TO} & 0 & 0\\
0 & 0 & \hbar\omega_{c}-|\epsilon_{z}| & \sqrt{2}iZ_{LO}\\
0 & 0 & -\sqrt{2}iZ_{LO} & \hbar\omega_{LO}\end{bmatrix}\begin{bmatrix}\Psi_{\uparrow\uparrow}\\
b_{+}\\
\Psi_{\downarrow\uparrow}\\
b_{z}\end{bmatrix}.\label{eq:Ham_nu1_gl0}\end{multline}
Therefore, at the crossing of the CR with the TO phonon when the electron
gas is fully spin-polarized, the Hamiltonian (\ref{eq:Ham_nu1_gl0})
gives the coupling parameter $\gamma=1$ which leads to the splitting
$\delta_{TO}=2Z_{TO}$, as shown in Fig. \ref{fig:TOsplitting}(a).
At the crossing of the CR with the LO phonon for this filling factor
and sign of the $g$-factor, the optically active mode is decoupled
from the phonons, giving $\delta_{LO}=0$. In the case where $g<0$,
described by the Hamiltonian in Eq. (\ref{eq:Ham_nu1_gl0}), the selection
rules represented in Fig. \ref{fig:JMS-table} show that the LO phonon
can couple with the optically inactive $\Psi_{\downarrow\uparrow}$
mode. The latter anticrossing is parameterised by $\gamma=2$ which
gives $\tilde{\delta}_{LO}^{s}=2\sqrt{2}Z_{LO}$, with the anticrossing
diagram plotted in Fig. \ref{fig:LOsplitting}(a).

When $g>0$ the ground state of the 2DEG with $\nu=1$ consists of
down-spin electrons. The excitations\cite{bib:noq} from this state
are $\Psi_{\downarrow\downarrow}$ and $\Psi_{\uparrow\downarrow}$
and the Hamiltonian is \begin{multline}
\hat{\mathcal{H}}_{\nu=1,g>0}=\left[\Psi_{\downarrow\downarrow}^{\dagger},\hat{b}_{+}^{\dagger},\Psi_{\uparrow\downarrow}^{\dagger},b_{z}^{\dagger}\right]\times\\
\times\begin{bmatrix}\hbar\omega_{c} & Z_{TO} & 0 & 0\\
Z_{TO} & \hbar\omega_{TO} & 0 & 0\\
0 & 0 & \hbar\omega_{c}-\epsilon_{z}^{\prime} & 0\\
0 & 0 & 0 & \hbar\omega_{LO}\end{bmatrix}\begin{bmatrix}\Psi_{\downarrow\downarrow}\\
b_{+}\\
\Psi_{\uparrow\downarrow}\\
b_{z}\end{bmatrix}.\end{multline}
Using the diagram in Fig. \ref{fig:JMS-table} we find that while
$\Psi_{\downarrow\downarrow}$ may couple to the $b_{+}$ phonon (which
leads to the same splitting $\delta_{TO}=2Z$ as for the $g<0$ case),
the conservation of $J_{z}$ forbids any coupling of $\Psi_{\uparrow\downarrow}$.
This means that there is no coupling between inter-Landau level excitons
and the $b_{z}$ phonon, so that $\tilde{\delta}_{LO}^{s}=0$.

The analysis above also applies to the case $\nu\ll1$ which represents
a very low density electron gas. 

In the 2DEG with $\nu=2$ both spin-up and spin-down states in the
lowest Landau level are filled so that the excitations in the 2DEG
with energy close to $\hbar\omega_{c}$ are $\Psi_{\uparrow\uparrow}$,
$\Psi_{\uparrow\downarrow}$, $\Psi_{\downarrow\uparrow}$ and $\Psi_{\downarrow\downarrow}$.
Figure \ref{fig:JMS-table} shows that linear combinations of $\Psi_{\uparrow\uparrow}$
and $\Psi_{\downarrow\downarrow}$ may couple to the $b_{+}$ phonon
and that $\Psi_{\downarrow\uparrow}$ may couple to the $b_{z}$ phonon.
The linear combinations which are relevant are the modes $S=(\Psi_{\uparrow\uparrow}+\Psi_{\downarrow\downarrow})/\sqrt{2}$
and $T=(\Psi_{\uparrow\uparrow}-\Psi_{\downarrow\downarrow})/\sqrt{2}$.
The mode $S$ is spinless and therefore corresponds to the CR. The
resulting Hamiltonian is\begin{multline*}
\hat{\mathcal{H}}_{\nu=2}=\left[S^{\dagger},T^{\dagger},b_{+}^{\dagger},\Psi_{\downarrow\uparrow}^{\dagger},b_{z}^{\dagger}\right]\times\\
\begin{bmatrix}\hbar\omega_{c} & 0 & 0 & 0 & 0\\
0 & \hbar\omega_{c} & \sqrt{2}Z_{TO} & 0 & 0\\
0 & \sqrt{2}Z_{TO} & \hbar\omega_{TO} & 0 & 0\\
0 & 0 & 0 & \hbar\omega_{c}+\epsilon_{z} & -\sqrt{2}iZ_{LO}\\
0 & 0 & 0 & \sqrt{2}iZ_{LO} & \hbar\omega_{LO}\end{bmatrix}\begin{bmatrix}S\\
T\\
b_{+}\\
\Psi_{\downarrow\uparrow}\\
b_{z}\end{bmatrix}.\end{multline*}
The uncoupled modes $b_{-}$ and $\Psi_{\uparrow\downarrow}$ are
not shown. This Hamiltonian reveals that for $\nu=2$ the CR is decoupled
from the phonons at the TO resonance because the spin polarisation
of the $\nu=2$ ground state is zero, that is $\delta_{LO}=0$. However,
coupling between the spinful mode $T$ and the TO phonon, and $\Psi_{\downarrow\uparrow}$
and the LO phonon are possible and are both parametrised by $\gamma=2$
leading to $\tilde{\delta}_{TO}^{0}=\tilde{\delta}_{LO}^{0}=2\sqrt{2}Z$.
The resulting anticrossing diagrams are shown in Fig. \ref{fig:TOsplitting}(b)
for the TO crossing, Fig. \ref{fig:LOsplitting}(b) for the LO crossing
with $g<0$, and Fig. \ref{fig:LOsplitting}(d) for the LO crossing
with $g>0$.

If the magnetic field is slightly tilted away from the $z$ direction
\textit{e.g.} as $\mathbf{B}=B_{\parallel}\mathbf{l}_{x}+B_{\perp}\mathbf{l}_{z}=B(\sin\theta\mathbf{l}_{x}+\cos\theta\mathbf{l}_{z})$,
then $J_{z}$ is no longer a good quantum number and additional couplings
become permitted. The Landau levels in the 2DEG are defined with respect
to the perpendicular component of the magnetic field and accordingly
$\lambda_{B}=\sqrt{\hbar/(eB_{\perp})}$. The coupling of the CR at
the TO resonance shows the same qualitative features as before, with
a reduced splitting value $\delta_{TO}=2Z_{TO}\cos\theta$ for $\nu=1$
whereas it remains as $\delta_{TO}=0$ for $\nu=2$. Regarding the
optically passive couplings near the TO resonance, the $s_{z}=0$
mode $T$ can couple to the $b_{+}$ phonon with $\tilde{\delta}_{TO}^{0}=2\sqrt{2}Z_{TO}\cos\theta$,
and the spinful modes $\Psi_{\uparrow\downarrow}$ and $\Psi_{\downarrow\uparrow}$
can each couple to the $b_{+}$ phonon with $\tilde{\delta}_{TO}^{s}=2Z_{TO}\sin\theta.$

Also the CR in a tilted field can show a filling-factor-dependent
coupling with the $b_{z}$ phonon which has the frequency $\omega_{LO}$.
At $\nu=1$ the splitting is given by $\delta_{LO}=\sqrt{2}Z_{LO}\sin\theta$
and at $\nu=2$ by $\delta_{LO}=0$. When $\nu=2$ the $T$ mode can
couple to the $b_{z}$ phonon with $\delta_{LO}^{0}=2Z_{LO}\sin\theta$,
and the $|s_{z}|=1$ modes can couple with the $b_{z}$ phonon with
$\tilde{\delta}_{LO}^{s}=\sqrt{2}Z_{LO}(\cos\theta+1)$ for the $\Psi_{\downarrow\uparrow}$
mode and $\tilde{\delta}_{LO}^{s}=\sqrt{2}Z_{LO}(\cos\theta-1)$ for
the $\Psi_{\uparrow\downarrow}$ mode.

In conclusion, we introduced an e-ph coupling mechanism in III-V semiconductor
quantum wells which is able to generate anticrossings between the
CR with and TO phonon mode, and the inter-Landau-level spinful magneto-exciton
with both the TO and LO phonons. The proposed e-ph interaction is
assisted by the spin-orbit coupling. The resulting anticrossings of
the electronic and lattice modes are dependent on the polarization
of the electron gas and therefore varies with the filling factor.
The coupling of the CR with the TO phonon is strongest when the electron
gas is fully spin-polarized - that is, at $\nu=1$ - and may be observable
experimentally if the CR is swept through the TO frequency for filling
factors in the range $0\leq\nu\leq1$. However, the effect is absent
in the vicinity of $\nu=2$. Tilting of the magnetic field also generates
a weak anticrossing between the CR and the LO phonon.

We would like to acknowledge helpful discussions with G. Martinez,
B. Narozhny and M. Goerbig. This work was funded by the Lancaster
Portfolio Partnership and ESF-FoNE project SpiCo.


\begin{thebibliography}{10}
\bibitem{bib:intham} The second quantised form of the electron-phonon
interaction Hamiltonian \eqref{eq:V_def} is \begin{multline*}
\hat{\mathcal{H}}_{e-ph}=\xi_{k}\sum_{n,n^{\prime},p,p^{\prime}}\int d\mathbf{x}\,\psi_{n,p}^{\ast}(\mathbf{p\times\hat{w}+\hat{w}\times p})\psi_{n^{\prime}p^{\prime}}\\
\times\begin{pmatrix}a_{n,p,\uparrow}^{\dagger} & a_{n,p,\downarrow}^{\dagger}\end{pmatrix}\boldsymbol{\sigma}\begin{pmatrix}a_{n^{\prime},p^{\prime},\uparrow}\\
a_{n^{\prime},p^{\prime},\downarrow}\end{pmatrix}\end{multline*}
 where $\xi$ is the matrix element of the $z$-dependent part of
the Hamiltonian.
\bibitem{bib:phonon} The density of states at the TO and LO frequencies
in the material which surrounds the quantum well is zero: these phonon
modes are confined and we use zero boundary conditions.
\bibitem{bib:noq} Note that the momentum index is missing from the
creation and annihilation operators since we only study excitations
with $q=0$.
\end{thebibliography}
\end{document}